\documentclass[fleqn,usenatbib]{mnras}

\usepackage[T1]{fontenc}

\DeclareRobustCommand{\VAN}[3]{#2}
\let\VANthebibliography\thebibliography
\def\thebibliography{\DeclareRobustCommand{\VAN}[3]{##3}\VANthebibliography}

\usepackage{graphicx}	
\usepackage{amsmath}	
\usepackage{amssymb}	
\usepackage{longtable}
\usepackage{lscape}
\usepackage{wasysym}
\usepackage{newtxtext,newtxmath}

\title[A target list for searching for habitable exomoons]{A target list for searching for habitable exomoons}

\author[V. Dobos et al.]{
Vera Dobos,$^{1,2}$\thanks{E-mail: vera.dobos@rug.nl}
Andr\'as Haris,$^{3,4}$
Inga E. E. Kamp,$^{1}$
and Floris F. S. van der Tak$^{1,5}$
\\
$^{1}$Kapteyn Astronomical Institute, University of Groningen, 9747 AD, Landleven 12, Groningen, The Netherlands\\
$^{2}$MTA-ELTE Exoplanet Research Group, 9700, Szent Imre h. u. 112, Szombathely, Hungary\\
$^{3}$Department of Physics, Faculty of Science, University of Helsinki, Yliopistonkatu 4, 00100 Helsinki, Finland\\
$^{4}$Konkoly Thege Mikl\'os Astronomical Institute, Research Centre for Astronomy and Earth Sciences, E\"otv\"os Lor\'and Research Network (ELKH),\\
1121, Konkoly Thege Mikl\'os \'ut 15-17, Budapest, Hungary\\
$^{5}$SRON Netherlands Institute for Space Research, Landleven 12, 9747 AD Groningen, The Netherlands
}

\date{Accepted XXX. Received YYY; in original form ZZZ}

\pubyear{2015}

\begin{document}
\label{firstpage}
\pagerange{\pageref{firstpage}--\pageref{lastpage}}
\maketitle

\begin{abstract}
We investigate the habitability of hypothethical moons orbiting known exoplanets. This study focuses on big, rocky exomoons that are capable of maintaining a significant atmosphere. To determine their habitability, we calculate the incident stellar radiation and the tidal heating flux arising in the moons as the two main contributors to the energy budget. We use the runaway greenhouse and the maximum greenhouse flux limits as a definition of habitability. For each exoplanet we run our calculations for plausible ranges of physical and orbital parameters for the moons and the planet using a Monte Carlo approach. We calculate the moon habitability probability for each planet which is the fraction of the investigated cases that lead to habitable conditions. Based on our results, we provide a target list for observations of known exoplanets of which the top 10 planets have more than 50\% chance for hosting habitable moons on stable orbits. Two especially promising candidates are Kepler-62~f and Kepler-16~b, both of them with known masses and radii. Our target list can help to detect the first habitable exomoon.

\end{abstract}




\section{Introduction}

Despite expectations that the Kepler or CoRoT space telescope would discover the first exomoons \citep{szabo06, simon07, kipping09b}, to date, there is no confirmed exomoon detection \citep[observations reported on moon candidates:][]{bennett14, kenworthy15, teachey18a, oza19, fox21, kipping22}. One obstacle could be their smaller size compared to planets, which makes their observation challenging. The smallest known exoplanets, however, are smaller than Ganymede and Titan \citep{barclay13, campante15}. We also know very massive giant planets, some exceeding 10 Jupiter-masses ($M_\mathrm{J}$). Could they have massive, Earth-sized moons that are massive enough to be detected with our current instrumentation?

Based on the examples seen in the Solar System, there seems to be a correlation between the mass of a planet and the mass of its regular satellites. The total mass of moons that were formed from the circumplanetary disc is approximately $10^{-4} M_\mathrm{p}$, where $M_\mathrm{p}$ is the mass of the planet \citep{canup04}. This means that a 10~$M_\mathrm{J}$ planet can have a 0.3~Earth-mass ($M_\oplus$) moon. Even though no exomoons are known, we have no reason to assume that moon formation does not occur the same way as it did in the Solar System. In fact, a circumplanetary disc was observed around PDS 70 c ($M_\mathrm{p} \approx$~1--10~$M_\mathrm{J}$, $M_\mathrm{disc} \approx$~0.7~$M_\oplus$), which supports the theory of regular moon formation in other planetary systems \citep{benisty21, portilla-revelo22}. Moons may also be formed by collision as in the case of our Moon \citep{hartmann75, cameron91, halliday00}. As a result of such a collision, larger moon-to-planet mass ratios are possible than in the case of regular moon formation from the circumplanetary disc. We do not know a maximum mass limit for moons being born by collision, but our Moon has $10^{-2} M_\oplus$. In this case, both the impactor and the Earth were rocky objects, and an impact with a gas giant may be qualitatively different \citep{barr16}. A study of \citet{malamud20} shows that in this case it is more likely that bigger moons form by the merging of several smaller moons which were formed not by one, but by several impacts. Another possible scenario is capturing moons which eliminates the question of moon formation. A local example for this is Triton \citep{mccord66, agnor06}. It was shown that by binary exchange capture, moons larger than $0.1 M_\oplus$ can be captured by giant planets \citep{williams13}. These are Mars-mass or heavier bodies. To date, 5 exoplanets are known with masses lower than that of Mars and 22 planets have masses lower than $1 M_\oplus$ according to The Extrasolar Planet Encyclopaedia\footnote{http://exoplanet.eu, data obtained on 2nd November 2021}. This implies that while detection of these small planets is challenging, we already have the technical capabilities of observing such small bodies. This is further supported by reports on exomoon candidates which are not confirmed for various reasons \citep[see the articles of][]{bennett14, kenworthy15, teachey18a, oza19, fox21, kipping22}. 

It seems that another obstacle for exomoon discoveries is the lack of stable orbits around close-in planets. The orbital evolution of moons is affected by tidal forces and as a result, a moon can escape from the planet (e.g. long-term evolution of the Moon), or a moon may get too close to the planet, reaching the Roche radius where it disintegrates (e.g. long-term evolution of Phobos). \citet{barnes02} have shown that moons on orbits around close-in planets are quick to get lost (in about 10$^4$--10$^5$ years), and the orbit stability increases with the distance from the star and with the mass of the moon. Several studies investigated this effect \citep[e.g.][]{alvarado17, zollinger17}, and a few recent studies applied it to known exoplanets \citep{guimaraes18, martinez-rodriguez19, tokadjian20, dobos21}. The two most successful exoplanet observation techniques (in terms of the number of new discoveries), the transit and the radial velocity technique, are biased towards detecting close-in planets, however, any moon around these planets can only stay in orbit for a short time (depending on the orbital and physical parameters, typically for less than a few million years). Consequently, planets in orbits of longer periods are more likely to have moons, but they are more difficult to observe \citep{szabo06, simon12}.

Even though no exomoons are known today, their potential habitability is an interesting and important question. Habitable moons with atmospheres can be observed in multiple ways (for example they are also good targets for spectroscopic observations) which can help their detection. Extending the search for life on exoplanets to including also exomoons expands our possibilities and chances to find habitable environments. Detecting the first habitable exomoons will provide great opportunities to study and characterize a new type of extra-solar bodies: exomoons.

The potential habitability of exomoons has already been studied in several works. If massive enough, there is no reason to assume that exomoons cannot provide a habitable environment. Similarly to the circumstellar habitable zone (HZ), a \textit{circumplanetary} habitable zone has been defined for Earth-like exomoons \citep[see e.g.][]{heller13, forgan14, heller14, forgan16, dobos17}. Different definitions exist, but most commonly the \textit{habitable edge} is used for its inner boundary which is defined by the runaway greenhouse effect; for the outer boundary, usually orbit stability considerations are taken into account. It is well known that beyond a certain fraction ($f$) of the planet's Hill sphere the moon escapes from the planet \citep[see e.g.][]{szebehely81, dvorak89, holman99, barnes02, donnison10}. However, there are different values used for this fraction, $f$, typically between 0.3 and 0.5. From the recent studies, the results of \citet{domingos06} are widely used. In this work we are applying their method for determining the maximum possible semi-major axes of moon orbits, which, besides $f$, takes into account also the eccentricity of the orbits of the planet and the moon (see Eq. 2 in Section~\ref{section:params} for prograde orbits).

Beside stellar radiation, tidal heating can be an important energy source for moons \citep{reynolds87, scharf06, heller13, peters13}. As a consequence, moons on eccentric orbits around planets that are outside the circumstellar habitable zone (meaning that their moons would be too cold to sustain liquid water on their surface), can still be habitable with sufficient tidal heating \citep{forgan14, dobos15, dobos17, zollinger17}. 

In this paper we are investigating the habitability of hypothetical moons that are on stable orbits around known exoplanets. The aim of this paper is to provide a target list 
to maximize the chance for detecting potentially habitable exomoons.
To determine their habitability, we are considering stellar radiation and tidal heating as the two main energy sources for moons, and thermal radiation and reflected light from the planet as minor contributors. We ignore radiogenic heating which may play a role in some cases, for example in maintaining a subsurface ocean in icy moons \citep[e.g.][]{tjoa20}.

\section{Methods}

In the following, we describe our method to determine the probability of having habitable moons on stable orbits around known exoplanets.

The planets were chosen from a catalogue, selected to include only those that cannot be brown dwarfs (i.e. with masses below 13~$M_\mathrm{J}$), which do not have brown dwarf hosts instead of stars (i.e. stellar mass should be above 0.08 $M_\odot$), and of which we know all the necessary parameters for our calculations. See the detailed description in Section~\ref{section:params}, which also lists all the parameter ranges used.

Applying a Monte Carlo approach, we simulated a hundred thousand moons around each planet: each of the physical and orbital parameters of both the planet and the generated moons were randomly selected 100~000 times, following specific distributions from plausible ranges (taking special case for parameters where the uncertainties are not symmetric; see Section~\ref{section:params}).

Then we calculated the incident stellar radiation, the tidal heating in the moon, the reflected light from the planet and the thermal energy from the planet. These four energy sources were considered in the calculation of the global heat flux on the moon. To determine whether a test moon is habitable, the global heat flux was compared to the \textit{runaway greenhouse} and \textit{maximum greenhouse} habitable zone boundary fluxes. This is what we use to define the \textit{habitability probability} of moons. See the details in Section~\ref{section:hab}.

Finally, in Section~\ref{section:validation}, we demonstrate that generating a hundred thousand test moons is enough to obtain statistically relevant results, because the habitability probability of moons around a given planet converges.

\subsection{Physical and orbital parameters} \label{section:params}

We made our calculations for the planets listed in the catalogue of The Extrasolar Planets Encyclopaedia\footnote{http://exoplanet.eu, data obtained on 30 September 2021}. Planets with the following parameters were excluded:
\begin{itemize}
    \item planets with masses above 13 $M_\mathrm{J}$, 
    \item planets with a host star below 0.08 $M_\odot$, 
    \item planets of which neither the orbital period ($P_\mathrm{p}$) nor the semimajor axis ($a_\mathrm{p}$) is known,
    \item planets for which none of the following three parameters are known: mass ($M_\mathrm{p}$), minimum mass ($M_\mathrm{p} \cdot \mathrm{sin} \, i$), and radius ($R_\mathrm{p}$).
\end{itemize} 
After this selection our list contained 4140 known planets.

To obtain the missing mass or radius of the planet, the Forecaster model \citep{chen17} was used. The uncertainties in the measured parameter were also taken into account when known. However, only symmetric errors were used as described in \citet[][Section 2.2]{chen17}. For these symmetric cases, the mean of the upper and lower errors was used as the standard deviation of a Gaussian distribution. Then a hundred thousand random values were generated for both the mass and radius (one of those from the Forecaster model, if needed) to obtain a representative distribution of realistic values for each planet. This way a different mass--radius pair was used in each of the hundred thousand runs for each planet, except for those cases where the uncertainty of the mass or radius parameter was not known or where it was not symmetric. In these cases the calculation was made without taking into account the uncertainties, i.e. using only a single value.

A Monte Carlo approach was used in our calculations; the following parameters were randomized within the given intervals for each of the 100~000 runs:
\begin{itemize}
    \item $e_\mathrm{m}$: eccentricity of the orbit of the moon; between 0.001 and 0.1 with a uniform distribution. This way the eccentricity covers two orders of magnitude, but is limited to values up to 0.1, because of the limitations of the tidal heating calculation method \citep{mignard80};
    \item $Q_\mathrm{p}$: tidal quality factor of the planet; depending on the type of the planet, the following distributions were used \citep[see][]{dobos21}:
        \begin{itemize}
        \item \textit{Rocky planets} ($R_\mathrm{p} < 2 R_\oplus$): $10 < Q_\mathrm{p} < 500$ with a log-uniform distribution;
        \item \textit{Ice/gas giants} ($R_\mathrm{p} \geq 2 R_\oplus$ and $P_\mathrm{p} > 10$~days): $10^3 < Q_\mathrm{p} < 10^6$ with a log-normal distribution with a mean of $\mu = 10^{4.5}$ and a standard deviation of $\sigma = 10^{0.5}$;
        \item \textit{Hot Jupiters} ($R_\mathrm{p} \geq 2 R_\oplus$ and $P_\mathrm{p} \leq 10$~days): a Gaussian distribution with $\mu = 5 \cdot 10^6$ and $\sigma = 2 \cdot 10^6$;
        \end{itemize}
    \item $M_\mathrm{m}$: mass of the moon; two constraints were applied:
        \begin{itemize}
        \item $0.01 M_\mathrm{p} < M_\mathrm{m} < 0.1 M_\mathrm{p}$ with a uniform distribution;
        \item 
        but never exceeding a maximum mass of 
        \begin{equation}\label{eq:Mm}
            M_\mathrm{m} \leq \frac {2} {13} \left( \frac {(f a_\mathrm{p})^3} {3 M_\star} \right)^{13/6}  \frac {M_\mathrm{p}^{8/3} Q_\mathrm{p}} {3 k_{2p} t R_\mathrm{p}^5 \sqrt{G}}
        \end{equation}
        \noindent to ensure that the test moons can stay on stable orbits for a long period of time, $t$ which is chosen to be the age of the system \citep{barnes02}. In the above equation 
        $M_\star$ is the mass of the host star and $G$ is the gravitational constant. The value of the Love number $k_{2p}$ was fixed to 0.5, which is not equally realistic for all types of planets, but since $k_{2p}$ only appears together with $Q$, it is enough to adjust the $Q$ parameter. This saves a free parameter in the calculations. The maximum mass of the moon was determined by using the lower value of the one calculated from Eq.~1 and $0.1 M_\mathrm{p}$;
        \end{itemize}
    \item $a_\mathrm{m}$: semi-major axis of the moon's orbit; between $2 R_\mathrm{p}$ and $a_\mathrm{max}$ with a log-uniform distribution to guarantee a stable (direct) orbit for the moon according to the criteria set by \citet{domingos06}:
        \begin{equation}
            a_\mathrm{max} = f (1 - 1.0305 e_\mathrm{p} - 0.2738 e_\mathrm{m}) R_\mathrm{Hill} \, ,
        \end{equation}
    \noindent where $f = 0.4895$ (fraction of the planet's Hill sphere beyond which the moon escapes). The same value was chosen for $f$ in Eq.~\ref{eq:Mm}. The Hill radius of the planet was calculated by
        \begin{equation}
            R_\mathrm{Hill} = a_\mathrm{p} \left( \frac {M_\mathrm{p}} {3 M_\star} \right)^{1/3} \, ;
        \end{equation}
    \item $\rho_\mathrm{m}$: bulk density of the moon; set depending on the distance from the snowline with a Gaussian distribution as described by \citet{dobos21}:
        \begin{itemize}
        \item if $a_\mathrm{p} < a_\mathrm{snow}$: $\mu = 3$~g~cm$^{-3}$ and $\sigma = 1/3$~g~cm$^{-3}$;
        \item if $a_\mathrm{snow} \leq a_\mathrm{p} < 2 a_\mathrm{snow}$: $\mu = 2.5$~g~cm$^{-3}$, $\sigma = 1/3$~g~cm$^{-3}$;
        \item if $2 a_\mathrm{snow} \leq a_\mathrm{p}$: $\mu = 1.5$~g~cm$^{-3}$, $\sigma = 1/6$~g~cm$^{-3}$;
        \item if $a_\mathrm{snow}$ is not known: $\mu = 2.5$~g~cm$^{-3}$, $\sigma = 1/2$~g~cm$^{-3}$;
    \end{itemize}
    where the location of the snowline around the star was estimated using the following equation \citep{cowan11}:
    \begin{equation}
        a_\mathrm{snow} = T_\mathrm{eff}^2 \cdot \frac {R_\star} {T_0^2} \, ,
    \end{equation}
    \noindent where $T_\mathrm{eff}$ is the effective temperature of the star and $T_0 \approx 230$~K is the equilibrium temperature at the planet’s sub-stellar point;
    \item $\alpha_\mathrm{p}$: Bond albedo of the planet; Gaussian distribution with $\mu = 0.3$ and $\sigma = 0.07$ to allow a wide range of possibilities, but with higher probabilities of moderate values;
    \item $\alpha_\mathrm{m}$: Bond albedo of the moon; Gaussian distribution with $\mu = 0.3$ and $\sigma = 0.07$, same as for the planet.
\end{itemize} 

The chosen $M_\mathrm{m}$ and $a_\mathrm{m}$ parameters described above ensure that the moon is on a stable orbit around the planet for a long period of time \citep[at least $10^4$ orbital periods of the planet,][]{domingos06}. The maximum mass for the moon was set with a conservative constraint, ignoring that in some cases even twice the mass would be allowed by the stability criterion \citep[see][]{barnes02}. If the age of the system ($t$) is not known, the age of the Solar System was used (4.57 billion years) for determining the possible largest mass of the moon. In general, large moons are considered in the calculations, because the aim of this work is to identify those planets which can host big, observable companions. Such large moons (between 0.01 and 0.1 $M_\mathrm{p}$) may not have been formed from a circumplanetary disc, but could have been captured like Triton, or formed by collision like our Moon ($0.0123 M_\oplus$) as described in the introduction.


If the generated moon radius is too large ($>2.5R_\oplus$) or if the moon mass is too low ($<0.1M_\oplus$), then this case is considered uninhabitable without applying any further habitability investigations. This is to avoid moons being mini-Neptunes instead of rocky moons, and also to avoid bodies that are not massive enough to keep a significant atmosphere.

%

\subsection{Habitability} \label{section:hab}

The global flux reaching the moon's surface is calculated as described by \citet{dobos17}:
\begin{equation}
\label{eq:fluxes}
  F_{\mathrm{glob}} = F_{*} + F_{\mathrm{ref}} + F_{\mathrm{th}} + F_{\mathrm{tidal}} \, ,
\end{equation}
where $F_{*}$, $F_{\mathrm{ref}}$, $F_{\mathrm{th}}$, and $F_{\mathrm{tidal}}$ are the stellar irradiation, the reflected light from the host planet, the thermal radiation of the host planet, and the tidal flux, respectively. The first three terms ($F_{*}$, $F_{\mathrm{ref}}$ and $F_{\mathrm{th}}$) were calculated following \citet[][Eq. 22]{heller13}. For the tidal heating flux, a viscoelastic model was used as described by \citet{dobos15}. They investigated the tidal heating of exomoons with a model based on the work of \citet{henning09} and \citet{moore03}. This model assumes a homogeneous rocky body with an inner convective and an outer conductive layer, using a Maxwell rheology. Tidal heating strongly depends on the viscosity and the shear modulus of the rocky material which, in turn, depends on the temperature of the body. These parameters are derived based on the equilibrium temperature at which tidal heating and convective cooling are in a stable equilibrium in the rock mantle.

An exomoon is considered habitable, if the global heating flux of the moon is between the runaway greenhouse ($F_\mathrm{RG}$) and maximum greenhouse fluxes ($F_\mathrm{MG}$). The runaway greenhouse flux was calculated with a method that is dependent on the surface gravity of the exomoon \citep[][Chapter 4]{pierrehumbert10}. 
The advantage of this method is that the runaway greenhouse flux scales with the radius and the mass of the body. 
This is in contrast to the standard calculation method described by \citet[][]{kopparapu14}, in which the calculations only apply to certain masses and radii.
It was shown, however, that the outer boundary of the circumstellar habitabile zone has a weak dependence on the mass of the exomoon \citep{kopparapu14}, and for this reason, the maximum greenhouse limit described by \citet{kopparapu14} was used as a lower limit for habitability.

We call the ratio of the habitable test moons compared to all cases (100~000) as \textit{habitability probability} and we measure it as a percentage. This parameter identifies those planets that have a high chance for hosting habitable moons on stable orbits.

\subsection{Validation of the number of test cases} \label{section:validation}

\begin{figure}
	\includegraphics[width=\columnwidth]{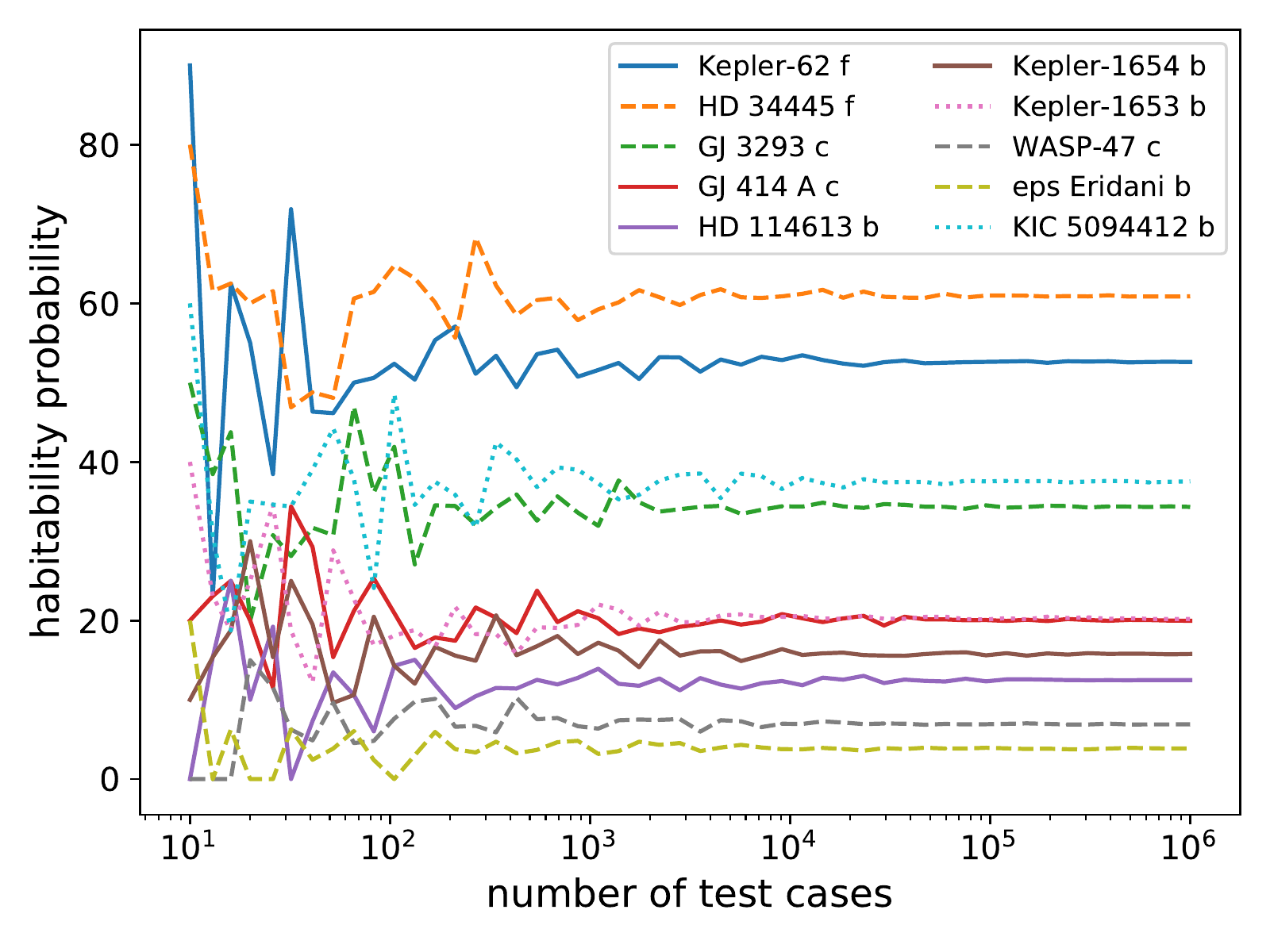}
    \caption{Habitability probability as a function of the number of test cases for ten planets. Solid curves: both mass and radius of the planet is known. Dashed curve: only the minimum mass of the planet is known. Dotted curve: only the radius of the planet is known.}
    \label{tests}
\end{figure}

To demonstrate that a hundred thousand test cases are enough for the habitability probability to converge, we show the habitability probability of a few selected planets as a function of the number of cases from ten to one million (see Fig.~\ref{tests}). The planets were chosen in such a way to show different habitability probabilities (ranging from low to high values) and also to represent planets with different measured data: transiting planets (only radius is known, dotted curves), radial velocity measurements (only minimum mass is known, dashed curves) and planets observed with both techniques (both mass and radius data available, solid curves). There seems to be no difference in the convergence for the planets; above a ten thousand runs the variation in the habitability probabilities is below 1.5~\% for each planet, and beyond a hundred thousand cases the curves are practically unchanging (the variations are below 0.35~\%).

\section{Results}

\begin{figure*}
	\includegraphics[width=15cm]{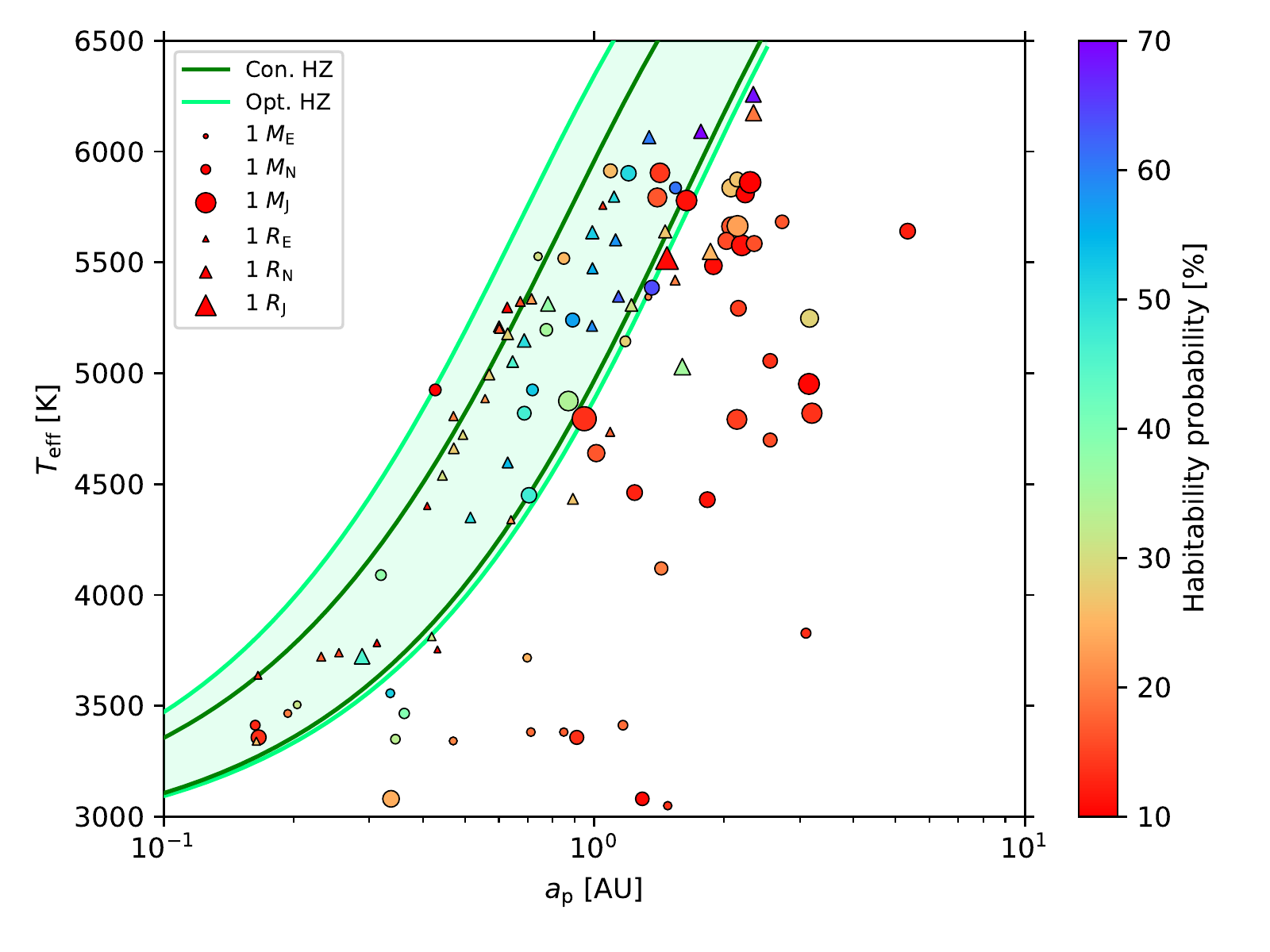}
    \caption{Habitability probability for exomoons around known exoplanets on the semi-major axis -- stellar effective temperature plane. Planets with known masses (with or without radius data) are marked with circles, planets with known radii only are marked with triangles. Colours of the markers correspond to the fraction of habitable moons and the sizes of the markers represent the sizes of the planets as shown in the legend. Note that the legend only shows three representative sizes (Earth, Neptune and Jupiter), while the size of the markers in the plot is scaled to the real size of the planets. Green curves represent the borders of the circumstellar habitable zone for a 1 Earth-mass planet: dark green for the consevative HZ (\textit{Con. HZ}) and light green for the optimistic HZ (\textit{Opt. HZ}).}
    \label{hzscatter}
\end{figure*}

After running the calculation for all 4140 exoplanets, a total of 234 planets were found with habitability probabilities $\geq 1~\%$ for moons. From these, 17 planets have a habitability probability higher than 50~\%. Fig.~\ref{hzscatter} shows exoplanets with at least 10\% habitability probability on the semi-major axis and stellar effective temperature plane. 
Planets with a measured mass (with or without radius data) are shown as circles, and planets with only radius data are shown as triangles. The area shaded in green represents the circumstellar HZ calculated for main sequence stars of different temperatures. To calculate the borders of this HZ, polynomial functions for a $1 M_\oplus$ planet given by \citet[][]{kopparapu14} were used together with stellar parameters (mass, temperature and luminosity) from the MIST database \citep{choi16} for 1~Gyr old (main sequence) stars
. A third order polynomial was fit to these discrete temperature values in order to obtain continuous curves for habitable zone limits. Along with the runaway greenhouse and maximum greenhouse limits (dark green curves representing the conservative HZ boundaries), a wider set of limits is also shown: the early Venus and the early Mars fluxes \citep[light green curves corresponding to the optimistic HZ boundaries,][]{kopparapu14}.

The boundaries of the circumstellar HZ are shown merely for illustration purposes and are calculated for objects with the same mass as the Earth. The runaway greenhouse flux for the moons, however, was calculated with the method of \citet[][]{pierrehumbert10}, that takes into account the surface gravitational acceleration of the moon.

\begin{figure*}
	\includegraphics[width=15cm]{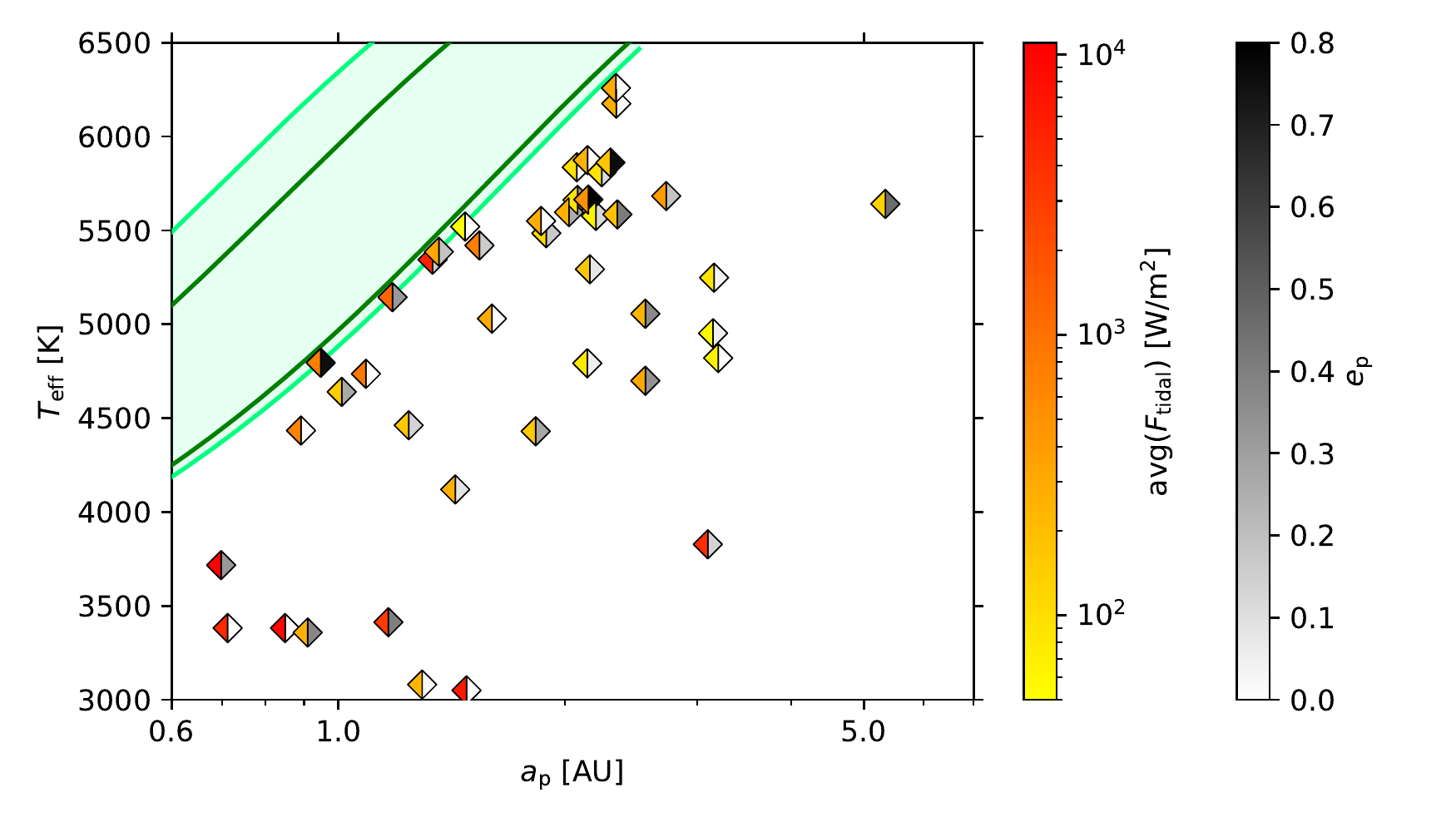}
    \caption{Average tidal fluxes of the simulated exomoons (yellow-to-red colours in the left hand side of the diamonds) and eccentricities of the orbits of host planets (white-to-black colours in the right hand side of the diamonds) as functions of the semi-major axis of the orbit of the planet and the effective temperature of the star. All planets shown in this figure are outside the optimistic HZ.}
    \label{eccscatter}
\end{figure*}

Conforming to expectations, a large number of planets with high habitability probability are found in the circumstellar HZ. This is because the stellar radiation alone (without additional heat sources for the moon) is already sufficient for supporting liquid water on the surface of an Earth-like body. 


Beyond the outer boundary of the HZ, where stellar radiation is weak and one would expect icy planets and moons, we still find a large number or planets with at least 10~\% habitability probability for moons. This is caused by the non-zero eccentricity of the orbit of the host planet (resulting in periodically experienced higher stellar fluxes) and also by the tidal heating arising in the moon. These two effects, if maintained on a long time-scale, can provide enough supplementary heat flux to prevent a global snowball phase of the moon (by pushing the flux above the maximum greenhouse limit). The contribution of these two factors to the total flux is presented in Fig.~\ref{eccscatter} for those planets which are outside the HZ. Two colour schemes correspond to each planet: one for the average tidal flux of the moons, $\mathrm{avg}(F_{\mathrm{tidal}})$ on a gradient from yellow to red, and the other one for the eccentricity of the orbit of the host planet, $e_{\mathrm{p}}$ indicated by a grey colour gradient. Note that the minimum average tidal heating flux in Fig.~\ref{eccscatter} is $57$~W/m$^{2}$ (yellow colour), which is significantly higher than the measured tidal heating flux on Io \citep[$\sim$2--4~W/m$^2$,][]{spencer00}. This means that tidal heating can be a significant contributor to the global energy of an exomoon, to the extent that it can make otherwise frozen environments habitable.

Planets with habitability probabilities above a reasonable but arbitrary limit of 10\% are shown in Table~\ref{tab:hab-frac}. Only those cases are presented for which the average exomoon radius of the 100~000 randomly simulated cases is larger than $1~R_\oplus$. 
Considering the size distribution of discovered exoplanets, a moon larger than Earth has a higher probability for a successful detection. Therefore we list
those planets, which 
can have
a detectable sized, potentially habitable moon. 
The planets are listed in order of decreasing habitability probability. The minimum, average, and maximum exomoon radii and masses used in the calculations are also shown. Planet mass and radius data are only shown if known from the catalogue.

The results for those cases where the average moon radius is equal to or smaller than $1~R_\oplus$ are shown in Table~\ref{tab:hab-frac-rest} in the Appendix.


An additional set of calculations was made with the same code only for those 49 planets that are shown in Table~\ref{tab:hab-frac}. This time only moons heavier than $1 M_\oplus$ were considered (as opposed to the minimum mass described in Section~\ref{section:params}) to see if the observable sized moons are indeed habitable
. For these calculations the number of test moons per planet was set to ten thousand. The results are shown in the Appendix in Table~\ref{tab:hab-cases}. For most planets the habitability probability did not change much. 
The results show, however, that in some cases when the planet is very massive, the moons generated in our simulation are often large enough to be mini-Neptunes, lowering the probability of habitable cases.

\section{Discussion}

We have investigated the habitability of hypothetical moons around known exoplanets. We have found that several massive planets have a high habitability probability for moons, regardless of their discovery method. Not surprisingly, planets with the highest habitability probability are in the circumstellar HZ. A high eccentricity of the planet's orbit and/or high tidal heating rates in the moons were found to be the main contributing factors of the habitability of exomoons outside the circumstellar habitable zone (where insufficient stellar radiation would make an Earth-like planet to be in a snowball state).

Although Fig.~\ref{hzscatter} shows the circumstellar HZ, the planets presented here were not filtered for main sequence host stars. The reason is that the spectral type of the star is 
not listed in our input catalogue for 2564 cases out of 4140 and even when it is known, its current evolutionary stage is not listed. This means that some of these planets can have host stars that are normally not considered to provide a stable radiation environment for life. This should be taken into consideration when selecting promising observation targets from our list. Note, however, that the giant planets themselves may provide extra protection as the moons are probably embedded in the planets' magnetospheres which can be strong enough to shield against stellar wind and galactic cosmic rays \citep{griessmeier09, hellerzulu13}. For smaller planets the coupled magnetospheres of the two bodies provide protection for both the planet and the moon \citep{green21}.

In the case of planet migration, it is likely that the giant planets in the circumstellar HZ were formed at larger distances from the star and then migrated inwards to their current orbit \citep[see for example][]{morbidelli10}. During the orbital migration they can lose some or all of their moons, especially if the moon orbit is close to the planet \citep{namouni10, spalding16}. Depending on the physical and orbital parameters of the planet and the moon, as well as on the starting and final semi-major axes of the planet, some moons can survive this process, and new moons can also be captured during or after the migration of the planet. Neither planet migration nor moon capture is included in our simulations as they require further studies, which are out of the scope of this article.

The provided lists of exoplanets (Table~\ref{tab:hab-frac} and Table~\ref{tab:hab-cases}) with high habitability rates for exomoons can be used as a target list for observations. Unfortunately, most of the planets with high habitability probability have orbital periods of several hundreds of days which makes observations of potential moons challenging. High on this list is Kepler-62~f with a 268~day orbital period, a measured mass and radius, and 53~\% habitability probability for moons. This planet has been the subject of several investigations as it is a 1.38~$R_\oplus$ planet in the circumstellar HZ of a K-type star \citep[see for example][]{borucki13, kane14, shan18}. An upper limit on its mass was placed at 35-36~$M_\oplus$ \citep{borucki13, borucki18}. \citet{sasaki14} studied the possibility of stable moon orbits around this planet and found that the lifetime for a moon would be longer than 5~Gyr in a large fraction of the tested configurations. These findings show that a large, Mars-, Venus- or maybe even Earth-sized moon (captured or formed by collision) can exist around this planet, and its orbit can be stable for a long timescale, providing a stable environment for life.

Another promising target that can be observed both by the transit and the radial velocity method is Kepler~16~(AB)~b, a 0.33~$M_\mathrm{J}$ planet on a 234.45~day orbital period around a double star. This was the first confirmed circumbinary planet which put the system in the center of attention \citep{doyle11}. The system consists of a K dwarf primary and an M dwarf secondary, accompanied by the approximately Saturn-sized planet, Kepler~16~b. \citet{forgan14b} found that the planet is outside the circumbinary habitable zone, while \citet{quarles12} shows that it is within the extended habitable zone which considers CO$_2$ clouds in the atmosphere allowing enhanced warming \citep{mischna00}. Orbital stability and climate models showed that if the planet has an Earth-sized moon or a Trojan companion, then it can be habitable and also observable \citep{quarles12, moorman19, sudol21}. This is in line with our result which shows a 47~\% habitability probability for a massive moon around the planet.

Many of the planets in Table~\ref{tab:hab-frac} are gas giants which makes it likely that they even have several moons. As discussed in the introduction, they can even have large, Mars- or Earth-sized moons. We propose to use our results to choose observation targets in the hope that they can lead to the discovery of the first habitable exomoon.

\onecolumn
\begin{table}
\begin{tabular}{lccccccccccc}
\hline
planet name & $M_\star$ & $M_\mathrm{p}$ & $R_\mathrm{p}$ & $P_\mathrm{p}$ & min($R_\mathrm{m}$) & avg($R_\mathrm{m}$) & max($R_\mathrm{m}$) & min($M_\mathrm{m}$) & avg($M_\mathrm{m}$) & max($M_\mathrm{m}$) & hab.prob. \\
 & [$M_\odot$] & [$M_\oplus$] & [$R_\oplus$] & [day] & [$R_\oplus$] & [$R_\oplus$] & [$R_\oplus$] & [$M_\oplus$] & [$M_\oplus$] & [$M_\oplus$] & [\%] \\
\hline
Kepler-459 b & 1.01 & -- & 5.38 & 854.08 & 0.54 & 1.35 & 2.50 & 0.10 & 1.64 & 10.23 & 70 \\
Kepler-456 b & 0.98 & -- & 6.46 & 1320.10 & 0.54 & 1.45 & 2.50 & 0.10 & 2.09 & 10.94 & 69 \\
HD 7199 b & 0.89 & 92.16 & -- & 614.11 & 1.06 & 1.76 & 2.50 & 0.76 & 4.11 & 11.26 & 64 \\
Kepler-1635 b & 0.89 & -- & 3.57 & 469.63 & 0.53 & 1.07 & 2.50 & 0.10 & 0.84 & 9.41 & 63 \\
HD 34445 f & 1.07 & 37.82 & -- & 676.84 & 0.63 & 1.52 & 2.46 & 0.16 & 2.08 & 6.49 & 61 \\
Kepler-458 b & 0.98 & -- & 4.50 & 572.38 & 0.52 & 1.23 & 2.50 & 0.10 & 1.24 & 9.49 & 60 \\
KIC 9662267 b & 0.86 & -- & 3.73 & 466.88 & 0.52 & 1.08 & 2.50 & 0.10 & 0.94 & 9.83 & 58 \\
Kepler-62 f & 0.69 & 34.96 & 1.38 & 267.54 & 0.79 & 1.48 & 2.17 & 0.35 & 1.92 & 3.50 & 53 \\
Kepler-47 (AB) c & 1.04 & -- & 4.50 & 351.79 & 0.52 & 1.23 & 2.50 & 0.10 & 1.24 & 9.94 & 52 \\
HD 564 b & 0.92 & 104.87 & -- & 500.62 & 1.06 & 1.59 & 2.50 & 0.79 & 3.81 & 11.28 & 51 \\
HD 137388 b & 0.86 & 70.87 & -- & 330.73 & 0.86 & 1.69 & 2.50 & 0.39 & 3.06 & 9.21 & 50 \\
Kepler-712 c & 0.84 & -- & 4.75 & 226.89 & 0.53 & 1.27 & 2.50 & 0.10 & 1.35 & 10.25 & 50 \\
Kepler-16 (AB) b & 0.85 & 105.83 & 8.27 & 234.45 & 1.14 & 1.57 & 2.50 & 0.86 & 3.77 & 11.13 & 47 \\
Kepler-1143 c & 0.81 & -- & 3.52 & 210.63 & 0.52 & 1.06 & 2.48 & 0.10 & 0.82 & 8.45 & 46 \\
KIC 10255705 b & 1.10 & -- & 7.13 & 704.88 & 0.53 & 1.02 & 2.50 & 0.10 & 1.44 & 11.09 & 36 \\
GJ 3293 c & 0.42 & 21.09 & -- & 122.64 & 0.65 & 1.14 & 1.90 & 0.17 & 0.91 & 2.41 & 34 \\
HD 218566 b & 0.85 & 66.74 & -- & 225.76 & 0.90 & 1.58 & 2.50 & 0.44 & 2.49 & 8.39 & 34 \\
55 Cnc f & 1.01 & 47.00 & -- & 246.56 & 0.87 & 1.57 & 2.44 & 0.47 & 2.33 & 4.70 & 34 \\
GJ 876 e & 0.33 & 15.43 & -- & 128.08 & 0.56 & 1.09 & 1.69 & 0.10 & 0.77 & 1.96 & 32 \\
Kepler-421 b & 0.79 & -- & 4.07 & 553.13 & 0.52 & 1.16 & 2.50 & 0.10 & 1.05 & 9.10 & 31 \\
HD 147379 b & 0.58 & 24.69 & -- & 86.54 & 0.70 & 1.14 & 1.91 & 0.25 & 0.94 & 2.47 & 28 \\
Kepler-967 c & 0.84 & -- & 3.58 & 198.71 & 0.52 & 1.07 & 2.49 & 0.10 & 0.85 & 9.44 & 28 \\
GJ 785 c & 0.78 & 24.15 & -- & 530.16 & 0.53 & 1.29 & 2.15 & 0.10 & 1.32 & 4.75 & 28 \\
Kepler-1628 b & 0.55 & -- & 6.30 & 76.38 & 0.54 & 1.21 & 2.50 & 0.10 & 1.34 & 10.23 & 28 \\
KIC 9704149 b & 0.86 & -- & 4.39 & 694.88 & 0.52 & 1.20 & 2.50 & 0.10 & 1.20 & 10.03 & 27 \\
HD 82943 d & 1.18 & 92.16 & -- & 1056.41 & 1.00 & 1.75 & 2.50 & 0.63 & 4.08 & 10.82 & 27 \\
Kepler-34 (AB)  b & 2.07 & 69.92 & 8.38 & 288.86 & 0.98 & 1.85 & 2.50 & 0.61 & 3.79 & 8.00 & 26 \\
Kepler-22 b & 0.97 & 35.91 & 2.33 & 290.14 & 0.79 & 1.49 & 2.25 & 0.36 & 1.98 & 3.59 & 25 \\
KIC 5010054 b & 1.05 & -- & 6.80 & 904.29 & 0.53 & 1.22 & 2.50 & 0.10 & 1.76 & 10.65 & 25 \\
HD 128356 b & 0.65 & 282.84 & -- & 367.67 & 1.51 & 1.42 & 2.50 & 2.00 & 3.20 & 11.35 & 21 \\
GJ 414 A c & 0.65 & 56.27 & 8.60 & 774.78 & 0.79 & 1.80 & 2.50 & 0.23 & 2.99 & 8.23 & 20 \\
Kepler-455 b & 0.98 & -- & 6.80 & 1322.30 & 0.55 & 1.45 & 2.50 & 0.10 & 2.17 & 10.50 & 19 \\
GJ 687 c & 0.41 & 15.89 & -- & 714.74 & 0.67 & 1.42 & 2.36 & 0.10 & 0.87 & 2.98 & 19 \\
GJ 273 d & 0.29 & 10.96 & -- & 407.52 & 0.68 & 1.27 & 1.88 & 0.11 & 0.60 & 1.10 & 19 \\
GJ 273 e & 0.29 & 9.44 & -- & 530.63 & 0.66 & 1.20 & 1.77 & 0.10 & 0.52 & 0.94 & 17 \\
HD 220197 b & 0.91 & 63.56 & -- & 1726.30 & 0.98 & 1.81 & 2.50 & 0.64 & 3.49 & 6.36 & 17 \\
HD 126614 A b & 1.15 & 120.76 & -- & 1229.79 & 1.11 & 1.38 & 2.50 & 0.83 & 3.34 & 11.41 & 16 \\
HD 219134 e & 0.79 & 70.87 & -- & 1679.12 & 1.06 & 1.84 & 2.50 & 0.71 & 3.51 & 7.09 & 16 \\
HD 164922 b & 0.87 & 116.00 & -- & 1240.38 & 1.17 & 1.42 & 2.50 & 1.16 & 3.44 & 11.54 & 15 \\
Proxima Centauri c & 0.12 & 9.53 & -- & 1898.60 & 0.65 & 1.20 & 1.86 & 0.10 & 0.52 & 1.52 & 14 \\
HD 204941 b & 0.74 & 84.53 & -- & 1739.30 & 0.98 & 1.62 & 2.50 & 0.46 & 3.22 & 9.42 & 14 \\
Lalande 21185 c & 0.46 & 18.11 & -- & 2939.65 & 0.80 & 1.50 & 2.28 & 0.18 & 1.00 & 1.81 & 14 \\
HIP 57050 c & 0.35 & 72.14 & -- & 532.96 & 1.25 & 1.17 & 2.50 & 0.72 & 1.40 & 5.87 & 13 \\
HD 113538 b & 0.58 & 114.41 & -- & 659.46 & 1.14 & 1.20 & 2.50 & 0.78 & 2.46 & 9.95 & 13 \\
HD 28254 A b & 1.06 & 368.64 & -- & 1118.50 & 1.70 & 1.33 & 2.50 & 3.69 & 3.30 & 11.39 & 13 \\
HD 114613 b & 1.27 & 113.45 & 12.97 & 3999.83 & 1.08 & 1.47 & 2.50 & 0.83 & 3.55 & 11.20 & 12 \\
WASP-107 c & 0.69 & 114.41 & -- & 1088.00 & 1.14 & 1.19 & 2.50 & 0.77 & 2.45 & 10.04 & 12 \\
GJ 3512 c & 0.12 & 63.56 & -- & 1529.58 & 1.21 & 1.28 & 2.50 & 0.64 & 1.55 & 5.86 & 11 \\
Kepler-62 e & 0.69 & 35.91 & 1.58 & 122.70 & 0.79 & 1.43 & 2.26 & 0.36 & 1.74 & 3.59 & 10 \\
\hline
\end{tabular}
\caption{List of planets with habitability probability (hab.prob., last column) above 10\%. Only those cases are shown where the average moon radius is above 1~$R_\oplus$. The mass of the star, the mass and radius of the planet, and the orbital period of the planet (columns 2--5, respectively) are from The Extrasolar Planet Encyclopaedia. The minimum, average, and maximum exomoon radii and masses generated for the calculations are shown in columns 6--11.}
\label{tab:hab-frac}
\end{table}
\twocolumn

\section*{Acknowledgements}

VD has been supported by the Hungarian National Research, Development, and Innovation Office (NKFIH) grant K-131508. The COFUND project oLife has received funding from the European Union's Horizon 2020 research and innovation programme under grant agreement No 847675.

\section*{Data Availability}

The data underlying this article are available in the article.



\bibliographystyle{mnras}
\bibliography{ref}




\appendix

\section{Planets with smaller test moons}

Table~\ref{tab:hab-frac-rest} lists the results of our calculations for those planets which have at least 10~\% habitability probability for moons (like in Table~\ref{tab:hab-frac}), but for which the average radius of the test moons is 1~$R_\oplus$ or smaller.

\onecolumn
\begin{table}
\begin{tabular}{lccccccccccc}
\hline
planet name & $M_\star$ & $M_\mathrm{p}$ & $R_\mathrm{p}$ & $P_\mathrm{p}$ & min($R_\mathrm{m}$) & avg($R_\mathrm{m}$) & max($R_\mathrm{m}$) & min($M_\mathrm{m}$) & avg($M_\mathrm{m}$) & max($M_\mathrm{m}$) & hab.prob. \\
 & [$M_\odot$] & [$M_\oplus$] & [$R_\oplus$] & [day] & [$R_\oplus$] & [$R_\oplus$] & [$R_\oplus$] & [$M_\oplus$] & [$M_\oplus$] & [$M_\oplus$] & [\%] \\
\hline
Kepler-1600 b & 0.86 & -- & 3.06 & 386.37 & 0.52 & 0.98 & 2.48 & 0.10 & 0.65 & 7.87 & 59 \\
Kepler-1634 b & 0.92 & -- & 3.13 & 374.88 & 0.51 & 0.99 & 2.50 & 0.10 & 0.67 & 8.49 & 56 \\
Kepler-1318 b & 0.73 & -- & 3.04 & 213.26 & 0.52 & 0.97 & 2.48 & 0.10 & 0.64 & 8.03 & 54 \\
Kepler-1636 b & 1.01 & -- & 3.16 & 425.48 & 0.52 & 1.00 & 2.49 & 0.10 & 0.69 & 7.62 & 50 \\
Kepler-1086 c & 0.70 & -- & 2.88 & 161.52 & 0.52 & 0.94 & 2.49 & 0.10 & 0.59 & 9.02 & 50 \\
GJ 752 A b & 0.45 & 12.20 & -- & 105.91 & 0.55 & 0.91 & 1.57 & 0.12 & 0.48 & 1.22 & 40 \\
KIC 5094412 b & 0.82 & -- & 5.60 & 277.89 & 0.53 & 0.91 & 2.50 & 0.10 & 1.16 & 11.27 & 38 \\
Kepler-1540 b & 0.74 & -- & 2.44 & 125.41 & 0.51 & 0.84 & 2.42 & 0.10 & 0.44 & 6.94 & 30 \\
Kepler-453 (AB) b & 0.93 & 9.53 & 6.04 & 240.50 & 0.52 & 0.95 & 2.10 & 0.10 & 0.68 & 4.84 & 29 \\
HD 11964 b & 1.12 & 197.67 & -- & 1934.58 & 1.32 & 0.77 & 2.50 & 1.37 & 1.95 & 11.45 & 29 \\
Kepler-1593 b & 0.81 & -- & 3.11 & 174.51 & 0.52 & 0.98 & 2.48 & 0.10 & 0.67 & 7.83 & 28 \\
Kepler-443 b & 0.74 & -- & 2.30 & 147.88 & 0.51 & 0.80 & 2.26 & 0.10 & 0.41 & 7.20 & 28 \\
Kepler-1341 b & 0.79 & -- & 2.93 & 133.00 & 0.52 & 0.95 & 2.48 & 0.10 & 0.60 & 7.47 & 27 \\
Kepler-1536 b & 0.71 & -- & 3.07 & 364.76 & 0.52 & 0.98 & 2.41 & 0.10 & 0.65 & 8.34 & 27 \\
HD 34445 b & 1.07 & 199.89 & -- & 1055.52 & 1.39 & 0.75 & 2.50 & 1.71 & 1.91 & 11.87 & 26 \\
GJ 3138 d & 0.68 & 10.49 & -- & 258.13 & 0.54 & 1.00 & 1.74 & 0.10 & 0.53 & 1.80 & 24 \\
Kepler-1690 b & 0.88 & -- & 2.75 & 234.81 & 0.52 & 0.91 & 2.49 & 0.10 & 0.55 & 8.51 & 23 \\
Kepler-441 b & 0.57 & -- & 1.65 & 247.72 & 0.52 & 0.53 & 2.07 & 0.10 & 0.21 & 4.35 & 21 \\
tau Cet f & 0.78 & 3.94 & -- & 636.04 & 0.51 & 0.63 & 1.04 & 0.10 & 0.20 & 0.39 & 21 \\
KIC 12454613 b & 0.87 & -- & 2.50 & 748.43 & 0.52 & 0.85 & 2.39 & 0.10 & 0.47 & 6.64 & 20 \\
Kepler-1653 b & 0.72 & -- & 2.13 & 138.98 & 0.51 & 0.76 & 2.10 & 0.10 & 0.36 & 4.78 & 20 \\
KOI-4427.01 & 0.53 & -- & 1.80 & 136.60 & 0.51 & 0.48 & 2.01 & 0.10 & 0.21 & 4.13 & 19 \\
Kepler-1544 b & 0.81 & -- & 1.74 & 168.81 & 0.52 & 0.47 & 1.90 & 0.10 & 0.19 & 3.72 & 18 \\
Kepler-1630 b & 0.66 & -- & 2.14 & 510.00 & 0.52 & 0.77 & 2.17 & 0.10 & 0.36 & 4.56 & 18 \\
HIP 57274 d & 0.73 & 167.48 & -- & 433.96 & 1.29 & 1.00 & 2.50 & 1.45 & 2.43 & 11.47 & 17 \\
HD 114729 A b & 0.93 & 266.95 & -- & 1136.28 & 1.56 & 0.52 & 2.50 & 2.67 & 1.38 & 11.37 & 17 \\
Kepler-68 d & 1.08 & 244.70 & -- & 582.25 & 1.48 & 0.59 & 2.50 & 2.23 & 1.52 & 11.90 & 17 \\
Kepler-1552 b & 0.85 & -- & 2.41 & 184.77 & 0.52 & 0.84 & 2.49 & 0.10 & 0.44 & 5.73 & 16 \\
Kepler-1654 b & 1.01 & 158.90 & 8.99 & 1048.17 & 1.27 & 0.99 & 2.50 & 1.59 & 2.46 & 11.60 & 16 \\
Kepler-705 b & 0.53 & -- & 2.06 & 56.06 & 0.52 & 0.58 & 2.22 & 0.10 & 0.27 & 5.64 & 16 \\
Kepler-1549 b & 0.88 & -- & 2.51 & 214.89 & 0.51 & 0.86 & 2.37 & 0.10 & 0.47 & 6.62 & 16 \\
7 CMA c & 1.52 & 287.92 & -- & 929.49 & 1.55 & 0.47 & 2.50 & 2.37 & 1.25 & 11.81 & 15 \\
Kepler-1097 b & 0.82 & -- & 3.19 & 187.75 & 0.53 & 1.00 & 2.48 & 0.10 & 0.70 & 8.35 & 14 \\
HD 17674 b & 0.98 & 276.48 & -- & 624.38 & 1.57 & 0.49 & 2.50 & 2.77 & 1.31 & 11.69 & 14 \\
Kepler-452 b & 1.04 & -- & 1.59 & 383.19 & 0.52 & 0.58 & 1.87 & 0.10 & 0.22 & 3.41 & 14 \\
HD 219415 b & 1.00 & 317.80 & -- & 2091.01 & 1.63 & 0.41 & 2.50 & 3.18 & 1.11 & 11.62 & 14 \\
GJ 357 d & 0.34 & 7.21 & -- & 57.55 & 0.51 & 0.55 & 1.36 & 0.10 & 0.19 & 1.21 & 14 \\
Wolf 1061 d & 0.25 & 7.69 & -- & 235.40 & 0.53 & 0.69 & 1.52 & 0.10 & 0.25 & 1.12 & 12 \\
Kepler-97 c & 0.94 & 343.22 & -- & 789.00 & 1.69 & 0.36 & 2.50 & 3.43 & 0.99 & 11.28 & 11 \\
HD 9174 b & 1.03 & 352.75 & -- & 1174.48 & 1.53 & 0.36 & 2.50 & 2.06 & 0.98 & 11.93 & 11 \\
HIP 14810 d & 0.99 & 181.14 & -- & 953.91 & 1.29 & 0.86 & 2.50 & 1.21 & 2.15 & 11.39 & 11 \\
KOI-771 b & 0.95 & -- & 13.50 & 670.65 & 0.66 & 0.29 & 2.50 & 0.16 & 0.69 & 11.09 & 11 \\
HD 73534 b & 1.29 & 365.47 & -- & 1798.05 & 1.72 & 0.33 & 2.50 & 3.65 & 0.91 & 11.30 & 11 \\
HD 170469 b & 1.14 & 212.92 & -- & 1146.97 & 1.43 & 0.70 & 2.50 & 2.13 & 1.79 & 11.57 & 11 \\
Kepler-1554 b & 0.84 & -- & 2.84 & 198.09 & 0.52 & 0.93 & 2.49 & 0.10 & 0.57 & 9.75 & 10 \\
\hline
\end{tabular}
\caption{List of planets similar to Table~\ref{tab:hab-frac}, but with test moons having an average radius, avg($R_\mathrm{m}$) $\leq 1 R_\oplus$.}
\label{tab:hab-frac-rest}
\end{table}
\twocolumn

\section{Planets with massive test moons}

Table~\ref{tab:hab-cases} shows the habitability probability for a second round of calculations which was made only for the 49 planets presented in Table~\ref{tab:hab-frac} (i.e. for those planets that had at least 10\% habitability probability for moons and where the average radius of the moons was at least 1~$R_\oplus$). This time, however, only moons having at least 1~$M_\oplus$ were considered in the calculations. The planets are listed in the same order as in Table~\ref{tab:hab-frac} to help comparing the data of the two tables. 

The ratio of cases when the moon is in runaway-greenhouse state (indicated as \textit{too hot cases}) or in a snowball state (\textit{too cold cases}) are also shown in Table~\ref{tab:hab-cases}, as well as the fraction of cases where the generated test moon was too massive to be considered as a rocky body. 

\begin{table}
\begin{tabular}{lcccc}
\hline
planet name & habitable & too hot & too cold & too massive \\
 & cases [\%] & cases [\%] & cases [\%] & cases [\%] \\
\hline
Kepler-459 b & 69.87 & 29.80 & 0.00 & 0.33 \\
Kepler-456 b & 68.54 & 28.36 & 0.00 & 3.10 \\
HD 7199 b & 64.70 & 24.61 & 0.00 & 10.69 \\
Kepler-1635 b & 63.22 & 36.72 & 0.00 & 0.06 \\
HD 34445 f & 61.06 & 38.93 & 0.00 & 0.01 \\
Kepler-458 b & 59.34 & 40.51 & 0.00 & 0.15 \\
KIC 9662267 b & 59.27 & 40.53 & 0.00 & 0.20 \\
Kepler-62 f & 52.34 & 47.66 & 0.00 & 0.00 \\
Kepler-47 (AB) c & 52.16 & 47.73 & 0.00 & 0.11 \\
HD 564 b & 50.15 & 28.82 & 0.00 & 21.03 \\
HD 137388 b & 50.82 & 37.84 & 0.00 & 11.34 \\
Kepler-712 c & 49.92 & 49.93 & 0.00 & 0.15 \\
Kepler-16 (AB) b & 48.08 & 30.39 & 0.00 & 21.53 \\
Kepler-1143 c & 46.56 & 53.42 & 0.00 & 0.02 \\
KIC 10255705 b & 36.05 & 36.81 & 0.00 & 27.14 \\
GJ 3293 c & 34.70 & 38.35 & 14.38 & 12.57 \\
HD 218566 b & 34.12 & 39.28 & 0.00 & 26.60 \\
55 Cnc f & 34.96 & 61.58 & 0.00 & 3.46 \\
GJ 876 e & 32.81 & 40.63 & 23.95 & 2.61 \\
Kepler-421 b & 32.53 & 33.39 & 34.05 & 0.03 \\
HD 147379 b & 28.23 & 47.04 & 0.00 & 24.73 \\
Kepler-967 c & 31.02 & 68.98 & 0.00 & 0.00 \\
GJ 785 c & 28.52 & 33.66 & 37.72 & 0.10 \\
Kepler-1628 b & 27.98 & 31.41 & 0.00 & 40.61 \\
KIC 9704149 b & 27.86 & 33.71 & 38.08 & 0.35 \\
HD 82943 d & 26.10 & 20.31 & 42.62 & 10.97 \\
Kepler-34 (AB)  b & 26.20 & 73.34 & 0.00 & 0.46 \\
Kepler-22 b & 26.47 & 73.53 & 0.00 & 0.00 \\
KIC 5010054 b & 27.12 & 56.58 & 0.00 & 16.30 \\
HD 128356 b & 22.49 & 22.55 & 0.00 & 54.96 \\
GJ 414 A c & 19.53 & 23.66 & 55.03 & 1.78 \\
Kepler-455 b & 19.19 & 23.50 & 52.24 & 5.07 \\
GJ 687 c & 19.11 & 31.46 & 49.41 & 0.02 \\
GJ 273 d & 18.23 & 31.67 & 50.10 & 0.00 \\
GJ 273 e & 17.45 & 30.53 & 52.02 & 0.00 \\
HD 220197 b & 17.46 & 21.17 & 61.28 & 0.09 \\
HD 126614 A b & 16.48 & 15.85 & 36.01 & 31.66 \\
HD 219134 e & 16.04 & 20.24 & 57.19 & 6.53 \\
HD 164922 b & 14.90 & 14.76 & 41.10 & 29.24 \\
Proxima Centauri c & 13.55 & 24.40 & 62.05 & 0.00 \\
HD 204941 b & 13.60 & 16.71 & 51.38 & 18.31 \\
Lalande 21185 c & 13.76 & 22.56 & 63.68 & 0.00 \\
HIP 57050 c & 12.90 & 14.17 & 28.88 & 44.05 \\
HD 113538 b & 12.90 & 12.91 & 32.93 & 41.26 \\
HD 28254 A b & 13.20 & 23.51 & 0.00 & 63.29 \\
HD 114613 b & 12.77 & 14.13 & 46.95 & 26.15 \\
WASP-107 c & 11.45 & 12.27 & 36.29 & 39.99 \\
GJ 3512 c & 9.94 & 12.00 & 38.85 & 39.21 \\
Kepler-62 e & 10.36 & 89.64 & 0.00 & 0.00 \\
\hline
\end{tabular}
\caption{Fraction of test moons that are habitable, in a runaway greenhouse, and in a snowball state in the second run of calculations (columns 2--4, respectively). Moon masses were limited to be at least 1~$M_\oplus$. The last column shows the fraction of test moons that were too big (mini-Neptunes) to be considered habitable.}
\label{tab:hab-cases}
\end{table}


\bsp	
\label{lastpage}
\end{document}